\documentclass[prl,twocolumn,prd,superscriptaddress,preprintnumbers,floatfix,showpacs,showkeys,letter]{revtex4-1}

\usepackage{url} 
\usepackage{lineno}
\usepackage{amsmath}

\usepackage{amscd}
\usepackage{amsmath}
\usepackage{amssymb}
\usepackage{graphicx}%

\usepackage[ruled]{algorithm2e}
\usepackage{needspace}
\usepackage{newtxtext}
\usepackage{booktabs}
\usepackage{bm,comment}

\usepackage[colorlinks=true, pdfstartview=FitV, linkcolor=blue,
            citecolor=blue, urlcolor=blue]{hyperref} 

\usepackage{sidecap}
\usepackage{grffile}

\usepackage{xcolor}
\usepackage{tikz}

\graphicspath{{./}}
\providecommand{\U}[1]{\protect\rule{.1in}{.1in}}
\newenvironment{proof}[1][Proof]{\noindent\textbf{#1.} }{\ \rule{0.5em}{0.5em}}

\definecolor{rred}{rgb}{0.7,0,0.1}
\definecolor{ccyan}{rgb}{0,.5,1}
\definecolor{greenrb}{rgb}{0.2,0.6,0.2}

\def\bea{\begin{equation} \begin{aligned}}
\def\eea{\end{aligned} \end{equation}}
\def\beas{\begin{equation*} \begin{aligned}}
\def\eeas{\end{aligned} \end{equation*}}
\def\bes{\begin{equation*}}
\def\ees{\end{equation*}}

\def\d{\, \mathrm{d}}

\def\be{\begin{equation}}
\def\ee{\end{equation}}
\def\adots{
  \mathinner{\mkern1mu\raise1pt\hbox{.}\mkern2mu\raise4pt\hbox{.}
  \mkern2mu\raise7pt\vbox{\kern7pt\hbox{.}}\mkern1mu}}

\def \xx{\bm{x}}

\def\R{\mathbf{R}}

\def\adots{
  \mathinner{\mkern1mu\raise1pt\hbox{.}\mkern2mu\raise4pt\hbox{.}
  \mkern2mu\raise7pt\vbox{\kern7pt\hbox{.}}\mkern1mu}}
  
\newtheorem{thm}{Theorem}[section]
\newtheorem{lem}{Lemma}[section]
\newtheorem{defi}{Definition}[section]

\newtheorem{rem}{Remark}[section]
\newtheorem{cor}{Corollary}[section]
\def\bt{\begin{thm}}
\def\et{\end{thm}}
\def\bl{\begin{lem}}
\def\el{\end{lem}}
\def\bd{\begin{defi}}
\def\ed{\end{defi}}
\def\bc{\begin{cor}}
\def\ec{\end{cor}}
\def\bp{\begin{proof}}
\def\ep{\end{proof}}
\def\br{\begin{rem}}
\def\er{\end{rem}}
\def\bi{\begin{itemize}}
\def\ei{\end{itemize}}

\def\W{{\boldsymbol{W}_t}}

\newcommand{\G}{\mathcal{G}}
\newcommand{\cL}{\mathcal{L}}
\newcommand{\FF}{\mathbf{F}}
\newcommand{\GG}{\mathbf{G}}
\newcommand{\GGG}{\mathcal{G}}

\begin{document}
\bibliographystyle{apsrev}

\title[Detecting and Attributing Change in Climate and Complex Systems]{Detecting and Attributing Change in Climate and Complex Systems: Foundations,  Green's Functions, and Nonlinear Fingerprints}

 \author{Valerio Lucarini}
\email{v.lucarini@leicester.ac.uk}
\affiliation{School of Computing and Mathematical Sciences, University of Leicester, Leicester, LE17RH, UK}
%
  \author{Micka\"el D. Chekroun}
\affiliation{Department of Earth and Planetary Sciences, Weizmann Institute of Science, Rehovot 76100, Israel}
 \affiliation{Department of Atmospheric and Oceanic Sciences, University of California, Los Angeles, CA}

\date{\today}

\begin{abstract} 
Detection and attribution (D\&A) studies are cornerstones of climate science, providing crucial evidence for policy decisions.
Their goal is to  link observed climate change patterns to anthropogenic and natural drivers via the optimal fingerprinting method (OFM).
We show that response theory for nonequilibrium systems offers the  physical and dynamical basis for OFM, including the concept of causality used for attribution. Our framework clarifies the method's assumptions, advantages, and potential weaknesses. We use our theory to perform D\&A for prototypical climate change experiments performed on an energy balance model and on a low-resolution coupled climate model. We also explain the  underpinnings of degenerate fingerprinting, which offers early warning indicators for tipping points.
Finally, we extend the OFM to the nonlinear response regime. Our analysis shows that OFM has broad applicability across diverse stochastic systems influenced by time-dependent forcings, with potential relevance to ecosystems, quantitative social sciences, and finance, among others. 
\end{abstract}




\maketitle
\subsection*{\normalsize Detection and Attribution of Climate Change}
\vspace{-.3cm}
\noindent The climate is a complex system comprising five subsystems---the atmosphere, the hydrosphere, the cryosphere, the biosphere, and the land surface---with a myriad of physical, biological and chemical processes interacting over a wide range of spatio-temporal scales  
\cite{Peixoto1992,Lucarini.ea.2014}.
 The resulting  dynamics is  multiscale, with different subsystems having a dominant role depending on the scales the climate is looked at \cite{ghil2008climate,neelin2010climate,stammer2018science,vonderHeydt2021}, with subtle interactions across scales that are yet to be understood \cite{liu2023opposing}. 
 The role of tipping points has emerged as a key climate change feature \cite{Lenton.tip.08,Boers2017,boers2021critical,Lohmann2021,Ditlevsen2023}, in particular due to the interplay of forcings, instabilities, and feedbacks \cite{Ghil2020,Rothman2017,liu2023opposing}.

The complexity of climate makes it difficult to disentangle climate variability and the forced climate change signal. 
Nonetheless, the  successive reports of the Intergovernmental Panel on Climate Change (IPCC) have confirmed two key scientific advances obtained via detection and attribution (D\&A) studies: 1)  statistical evidence of the change of the current state of the climate system with respect to the conditions prevailing in the XIX and early XX century; and 2) statistical evidence that such a change can be attributed (for the most part) to anthropogenic causes \cite{IPCC13,IPCC2021}. More recently, D\&A studies have also investigated climate change at a local level \cite{Stott2010} including  the relationship between climate change and extreme events \cite{Easterling2016}. The current focus on extremes in a changing climate \cite{Ghil2011,IPCC12} and on tipping points has recently led the scientific community to use expressions like \textit{climate crisis} or \textit{climate emergency} instead of \textit{climate change} \cite{Ripple2021,Rodgers2023}.

D\&A studies wish to link in a causal sense the observed climate change  with various  acting forcings  \cite{Hasselmann1993b,Hegerl1996,Hasselmann1997}. 
The Optimal Fingerprinting Method (OFM) \cite{Allen1999,allen2003estimating,IPCC13,Hannart2014} aims  
at expressing a (vector-valued) observed climate change signal $Y_k$, $k=1,\ldots,S$ as a linear combination of response patterns: 
\begin{equation} \begin{aligned}
\label{eq:da3}
Y_k=\sum_{p=1}^{M}\tilde{X}_k^p\beta_p+\mathcal{R}_k, \hspace{.15cm}
\tilde{X}_k^p={X}_k^p+\mathcal{Q}_k^{p},\hspace{.15cm} k=1,\ldots S
\end{aligned}\end{equation} 
where $\tilde{X}_k^p$'s are the $M$---typically $\leq\mathcal{O}(10)$---fingerprints, each associated with one forcing.
One seeks  the optimal solution in the $\beta_p$'s  to the (strongly underdetermined, $S\gg M$ in all cases of interest) multivariate regression problem above, where the $\mathcal{Q}_k^{p}$'s account for sampling or model uncertainties, and the residual $\mathcal{R}_k$ is the natural climate variability \cite{Hasselmann1993b,Hegerl1996,Hasselmann1997,Allen1999,allen2003estimating,Hannart2014}.
Usually $\mathcal{R}_k$ and $\mathcal{Q}_k^p$ are modelled as independent, normally distributed stochastic vectors with zero mean and given covariance matrices. 
The estimate $X_k^p$  for the $p^{th}$ fingerprint 
is  obtained by averaging across an ensemble of forced model runs, each driven by the same protocol for the $p^{th}$ forcing, whilst the other forcings are switched off \cite{stone2007detection,hegerl2011use}. 

An example follows. Let $Y_k$ be a gridded surface temperature ($T_S$) anomaly field measured at $S$ locations. Assuming that the $p^{th}$ forcing is the increase in the $CO_2$ concentration ($[CO_2]$), $\tilde{X}_k^p$ features larger anomalies  at high latitudes as a result of polar amplification. If the $q^{th}$ forcing is a localized increase in the (reflecting) aerosol concentration, then $\tilde{X}_k^q$ is strongly spatially heterogeneous with negative values where such concentration is higher.

D\&A relies on Pearl causality \cite{Pearl2009}. This angle  allows for estimating causal effects through an interventionist protocol \cite{Hannart2016,wuebbles2017climate}. Each fingerprint is computed using climate models - which provide the counterfactual reality -  where the intervention is the selective application of one  forcing. Observations-only-based D\&A requires strong assumptions on the link between climate change and  climate variability \cite{hegerl2011use}.

The attribution of the  signal to the $p^{th}$ forcing depends on the $\beta_p$-confidence interval. If this  includes $1$ and excludes $0$, we can  reject the null hypothesis of no influence. The narrower this interval, the more robust the D\&A outcome. One has a favorable signal-to-noise ratio if either the   signal is highly sensitive to the  forcing, or if the natural variability is weak.  Attribution is also impacted by model uncertainties in  estimating the  fingerprints. Strong  interactions between forcings can significantly skew the $\beta$'s estimate. Finally, overlooking  key forcings can  jeopardise  D\&A, leading to erroneous  attribution. 

\vspace{-.3cm}
\subsection*{\normalsize This Work }
\vspace{-.3cm}
\noindent 
Traditionally, OFM relied on an empirical justification based on Eq.~\eqref{eq:da3}. We demonstrate that Eq.~\eqref{eq:da3} can be rigorously derived from linear response theory (LRT) applied to non-equilibrium systems, a well-established framework in statistical mechanics for both deterministic \cite{ruellegeneral1998,ruelle2009}  and stochastic systems \cite{Hairer2010}. The Green's function formalism plays a key role here in linking cause (forcing) and effect (observed signal).

This work offers a three-fold contribution to climate physics. First, we establish a robust physical and dynamical foundation for OFM. This foundation allows for more advanced fingerprinting analysis, facilitating a deeper understanding of cause-and-effect relationships within the climate system. Second, our derivation provides a theoretical framework for the degenerate fingerprinting method used for tipping point detection, a crucial component of early warning systems  \cite{Held2004,Lenton2012,Boettner2022}. Third, by leveraging nonlinear response theory \cite{ruelle_nonequilibrium_1998,lucarini2009b}, we  propose an extension of OFM accounting for the nonlinear effects tied to the interplay between multiple climate forcings.

To demonstrate the real-world applicability of our theoretical advancements, we conduct detailed analyses on two climate models with varying complexity. This includes the one-dimensional Ghil-Sellers (G-S) energy balance model \cite{Ghil1976} and the open-source, low-resolution coupled climate model PLASIM \cite{Fraedrich2005,Lucarini2010a,Lucarini2010b,Mehling2023}.

\vspace{-.3cm}
\subsection*{\normalsize Stochastic Climate Dynamics and Response Theory}
\vspace{-.3cm}
\noindent Following Hasselmann's program \cite{hasselmann1976,Imkeller2001}---see also the recent review  \cite{LC2023}---we consider climate models formulated as stochastic differential equations (SDEs), where the  impact of unresolved scales on the scales of interest has been stochastically parameterized \cite{majda2001mathematical,harlim2012,MSM2015,Berner2017,santos2021reduced}.  We study climate change by looking at how perturbations acting on the dynamics affect  the ensemble statistics. Hence, the 
unperturbed dynamics is described by the following $d$-dimensional It\^o SDE \cite{vanKampen1981,pavliotisbook2014}, 
\be\label{Eq_unpertSDE}
\d \xx =\FF(\xx) \d t +\Sigma(\xx) \d \W.
\ee
Here, $\FF$ (drift  term) is a smooth vector field  on $\mathbb{R}^d$, $\W$ denotes a $p$-dimensional Wiener process ($p\geq 1$), $\Sigma$ is the diffusion coefficient matrix (of size $d\times p$),  and $\xx$ describes the climate state.  The choice of the climate subcomponents included in our model and of the explicitly resolved spatial scales  impacts the drift and noise laws   
\cite{hasselmann1976,Imkeller2001,majda2001mathematical,Kondrashov_al2018_QG,saltzman_dynamical,LC2023}.  Correspondingly, $d$ can range from $\mathcal{O}(10^1)$ to $\mathcal{O}(10^8)$ \cite{Ghil2020}; yet our framework can be seamlessly applied. 

We consider here time-dependent forcings impacting the resolved scales; see  \cite{Santos2022} for a more general case. We study: 
\be\label{eq:sto ode 2}
\d \xx  = \left(\FF (\xx)  + \sum_{p=1}^M\epsilon_p g_p (t)\GG_p(\xx) \right)\d t+\Sigma(\xx)\d \W, 
\ee
where the ``large-scale'' perturbations 
are embodied by one or more ``pattern forcing'' $\GG_p$ of $ \R^d$, each modulated by a scalar-valued bounded function $g_p(t)$ and by a parameter $\epsilon_p \ll1$.

The climatology of any smooth observable $\Psi$ is the expected value under the reference climate: $\langle \Psi \rangle_0= \int \d\xx\rho_0(\xx)\Psi(\xx)$. Here, $\rho_0(\xx)$ represents the stationary probability density of the unperturbed system (Eq.~\eqref{Eq_unpertSDE}), solving its associated Fokker-Planck equation (FPE) \cite{risken,pavliotisbook2014}. Good choices for $\Psi$ include essential climate variables (ECVs) \cite{Bojinski2014}, which are key physico-chemical quantities observed at different spatial scales, as well as metrics used to evaluate Earth System Models (ESMs) \cite{Eyring2020}.

Response theory allows one to predict the system's response to perturbations from properties of the unperturbed system, as exemplified by the 
celebrated fluctuation-dissipation theorem (FDT) \cite{Kubo1966,marconi2008fluctuation}. 
In a seminal article, Leith \cite{Leith75} suggested that if FDT can be applied in 
climate science, then climate change projections can be performed using the statistical properties of the natural climate variability. 
Linear response theory (LRT) for nonequlibrium systems \cite{ruellegeneral1998,ruelle2009,Hairer2010} indicates that  
the first-order correction  $\delta^{(1)}[\Psi] (t) $ to the statistics  $\langle \Psi \rangle_0$  is given by a suitable 
lagged correlation between functions of the unperturbed system \cite{majda2005information}. 
The Green's function formalism enables for useful insights as discussed below. 
It states that 
\begin{equation} \begin{aligned}\label{eq:linear response time dependent}
\delta^{(1)}[\Psi] (t) = \sum_{p=1}^M\epsilon_p \int_{-\infty}^t  \d s\GGG^p_{\Psi} (t-s) g_p(s),
\end{aligned}
\end{equation} 
where the $\GGG^p_{\Psi}$ are the (causal) Green's functions given in Appendix A. 
The quality of approximation of the actual change in statistics by  $\delta^{(1)}[\Psi] (t) $ depends on many factors. 
It deteriorates with strong perturbations and/or if the unperturbed system has  slow decay of correlations \cite{ruelle_nonequilibrium_1998,lucarini2009b,Lucarini2016,SantosJSP,Santos2022}, such as encountered when approaching tipping points \cite{Santos2022}. 
 \vspace{-.3cm}
\subsection*{\normalsize Connecting OFM to LRT via Green's Functions}
\vspace{-.3cm} 
\noindent We now show how the structural equation of OFM---Eq.~\eqref{eq:da3}---can be derived from LRT.
Let $\Psi_k$ be a climatic observable with reference climatology $\langle \Psi_k \rangle_0$. 
We want to assess the contribution of the large-scale forcing, $g_p(t) \GG_p(\xx)$, to the anomaly signal $Y_k (t)=\Psi_k(t)-\langle \Psi_k\rangle_0$. Trivially,
\begin{equation} \begin{aligned}\label{Eq_fluc_ensemble}
Y_k(t)=\Psi_k(t) -\langle \Psi_k \rangle_{\rho_\varepsilon^t}+\langle \Psi_k \rangle_{\rho_\varepsilon^t} - \langle \Psi_k\rangle_0,
\end{aligned}\end{equation} 
 where $\langle \Psi_k \rangle_{\rho_\varepsilon^t}$ is the ensemble average with respect to $\rho_\epsilon^t$, which evolves according to the FPE associated with Eq.~\eqref{eq:sto ode 2}. 
The difference term between the ensemble averages in Eq.~\eqref{Eq_fluc_ensemble} is the focus of response theory. By approximating it by Eq.~\eqref{eq:linear response time dependent}, we obtain:
\bea\label{eq:da2}
Y_k(t)&=\sum_{p=1}^M\tilde{X}_k^p(t) +\mathcal{R}_k(t), \quad k=1,\ldots,N,\\
&\mbox{with } \tilde{X}_k^p(t)=\epsilon_p \int_{-\infty}^t  \d s \GGG^p_{\Psi_k} (t-s) g_p(s).
\eea 
The terms $\tilde{X}_k^p(t)$
account for the forced variability, whilst
$\mathcal{R}_k(t)=\Psi_k(t) -\langle \Psi \rangle_{\rho_\varepsilon^t}$ is a stochastic vector whose correlations are described by the  statistical state  $\rho_\varepsilon^t$. 
When $\Sigma=0$, this statistical state is carried by the system's pullback attractor at time $t$ \cite{Chekroun2011,CGN17,Chekroun_al22SciAdv} also known as the snapshot attractor \cite{Tel2020}.
 
 Let us compare Eqns.~\eqref{eq:da2} and \eqref{eq:da3}. The fingerprints $\tilde{X}_k^p(t)$ are constructed as convolution products of the Green's function relative to the considered forcing with the corresponding time modulation. Hence, it makes perfect sense to approximate them---as customarily done in most D\&A analyses---as ensemble average of the response signals obtained by applying selectively the corresponding forcing. This  mirrors an efficient method used for estimating the Green's functions  \cite{Ragone2016,Lucarini2017,Lembo2020}. 
 
In  Eq.~\eqref{eq:da2} there is a conspicuous absence of the $\beta_p$'s. This is actually a key element of our derivation:  LRT predicts 
that all the $\beta_p$'s should be unitary, apart from  (marginal) uncertainty, when D\&A is performed with a single ``perfect'' model.
 
\vspace{-0.5cm}
\subsection*{\normalsize Detection and Attribution of Climate Change: Examples}
\vspace{-0.3cm}
 
\begin{figure*}[htbp]
\centering
\includegraphics[width=0.24\linewidth, height=.14\textwidth]{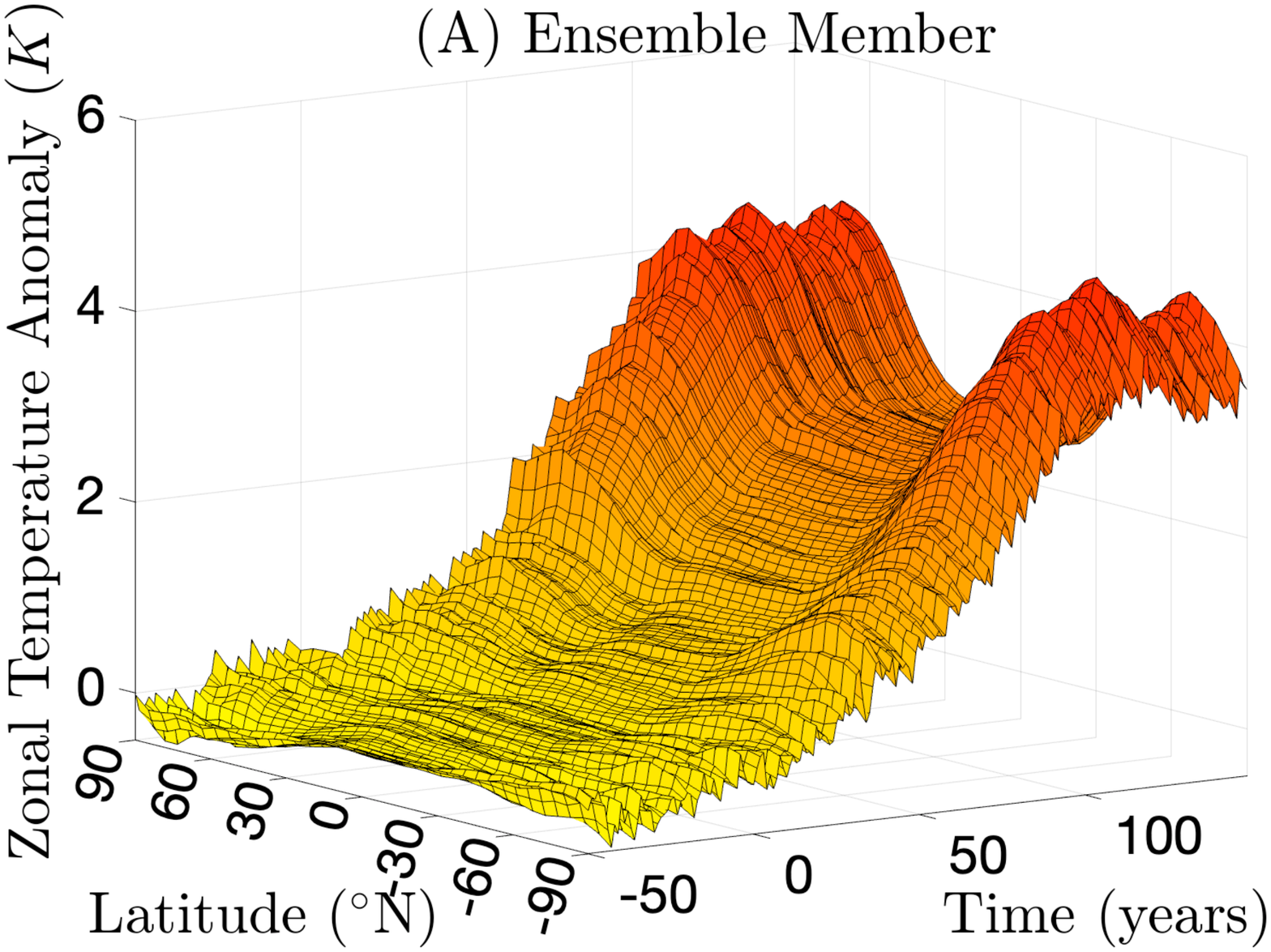}
\includegraphics[width=0.24\linewidth, height=.14\textwidth]{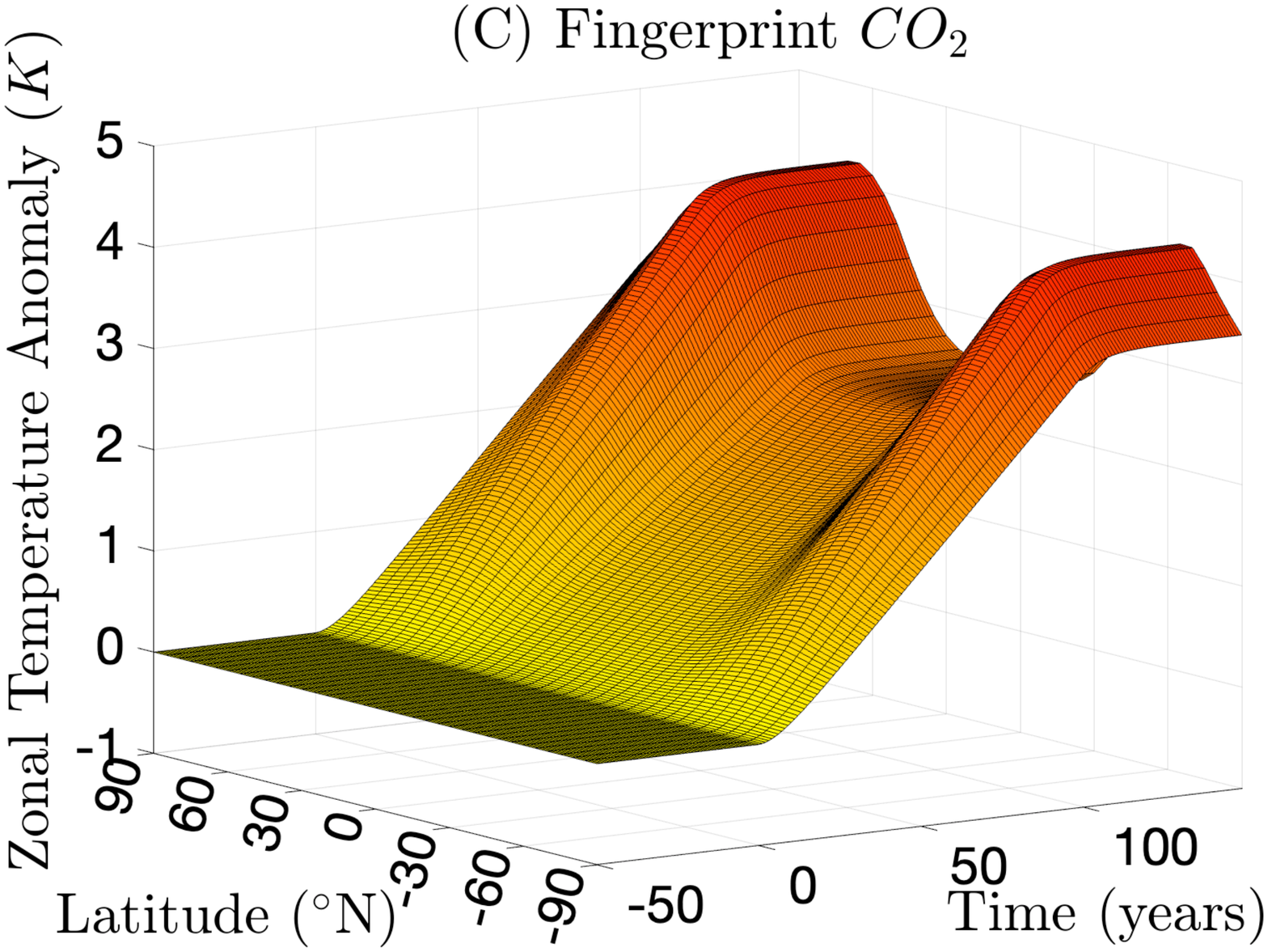}
\includegraphics[width=0.24\linewidth, height=.14\textwidth]{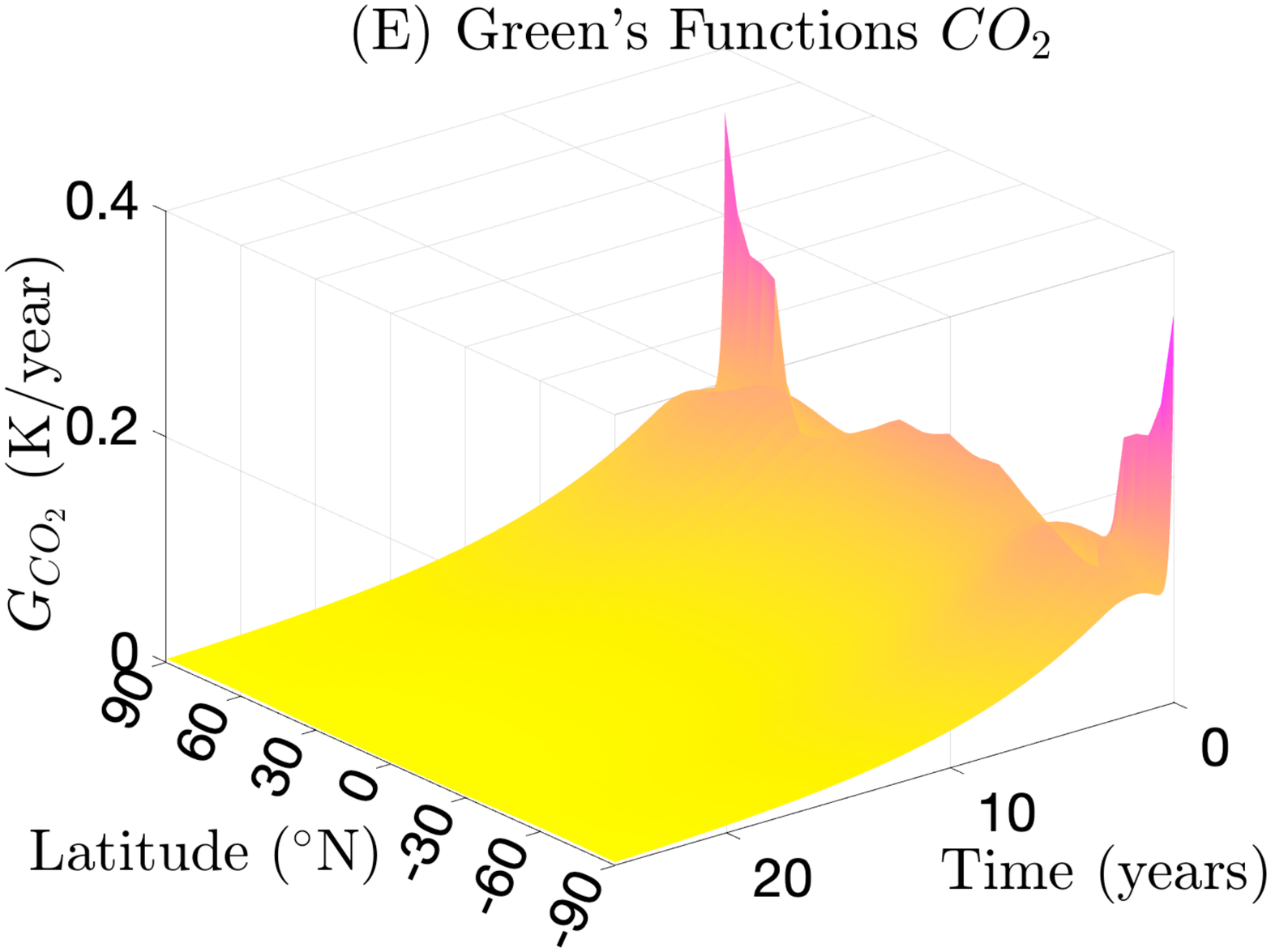}
\includegraphics[width=0.24\linewidth, height=.14\textwidth]{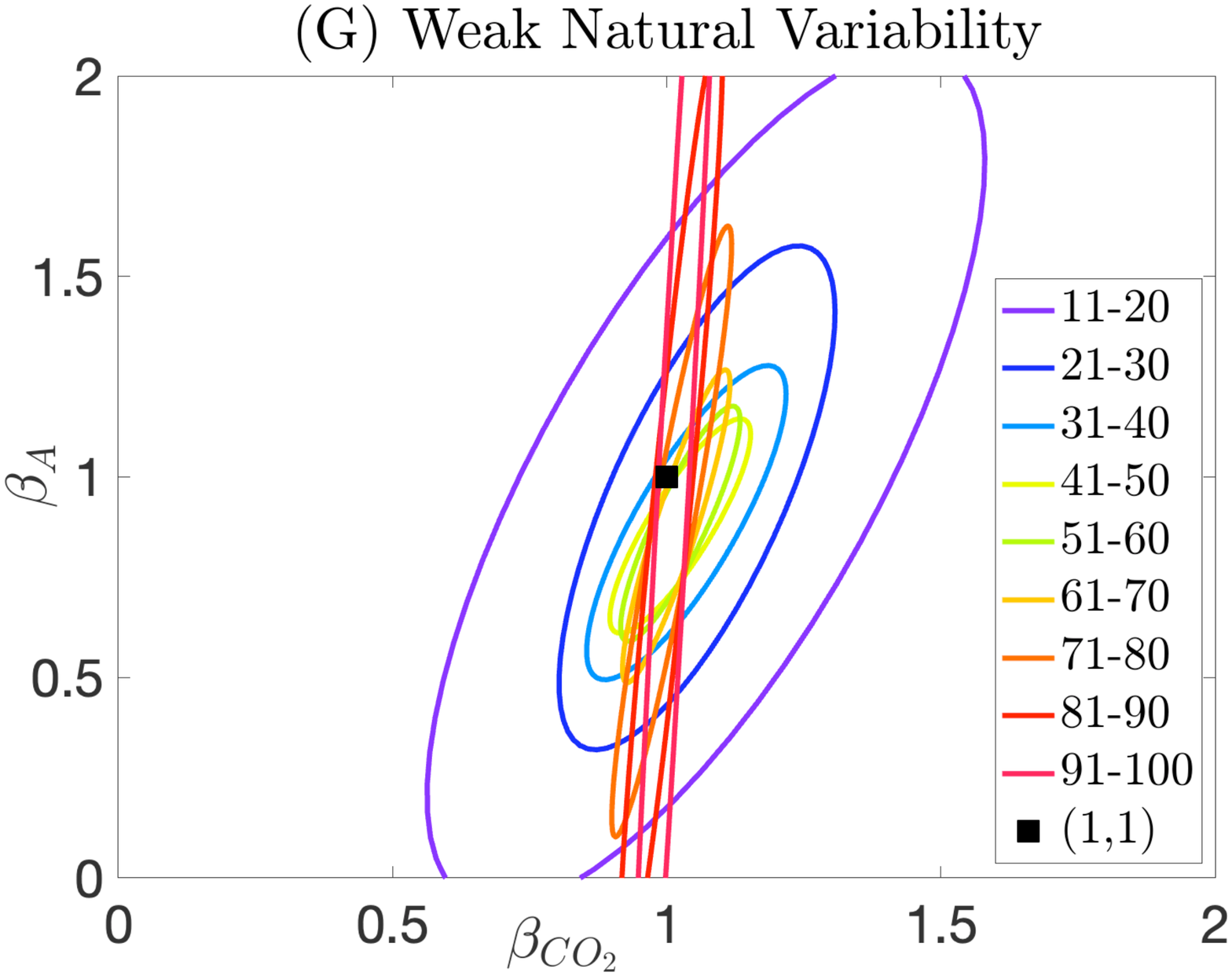}\\
\includegraphics[width=0.24\linewidth, height=.14\textwidth]{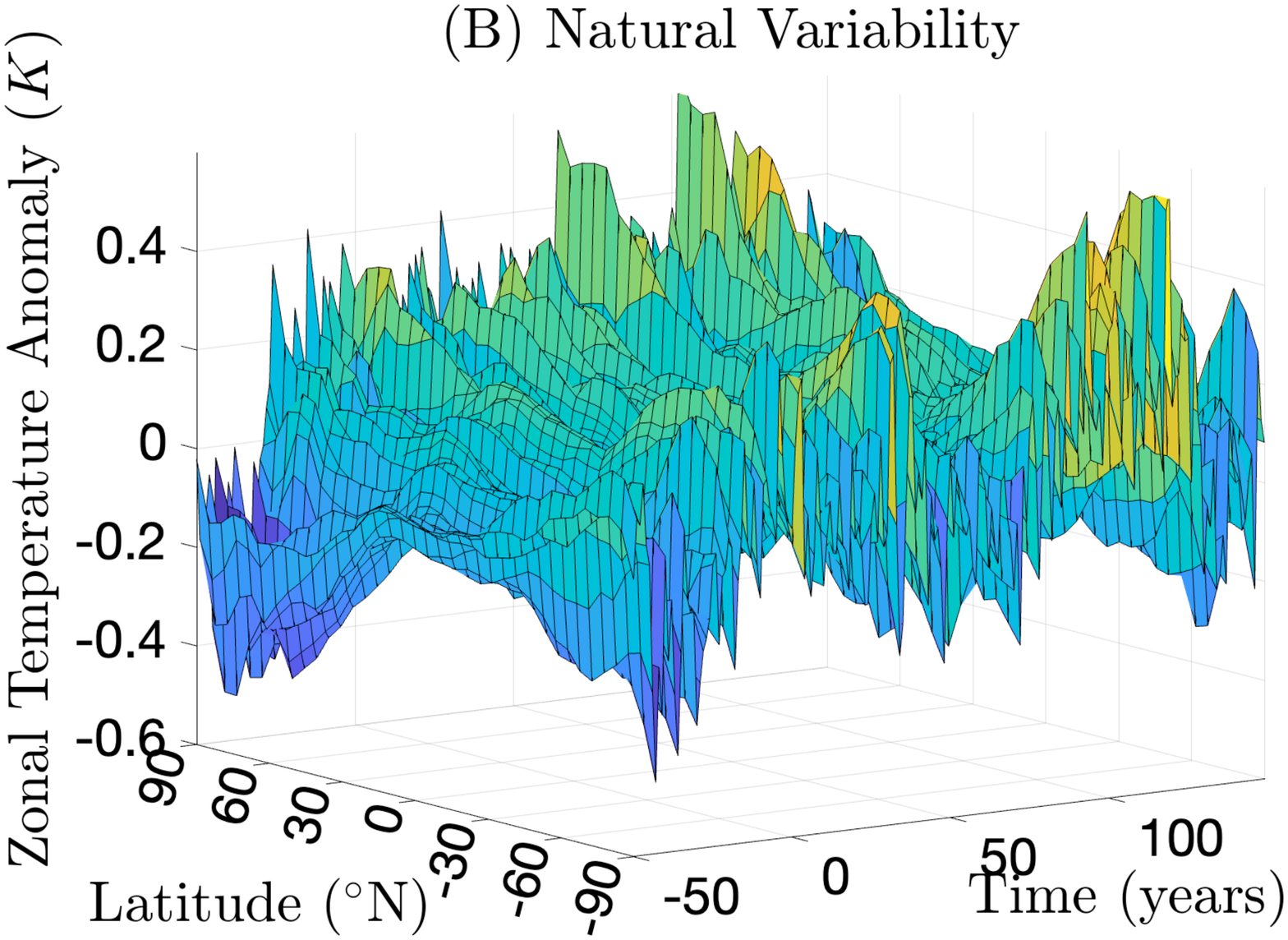}
\includegraphics[width=0.24\linewidth, height=.14\textwidth]{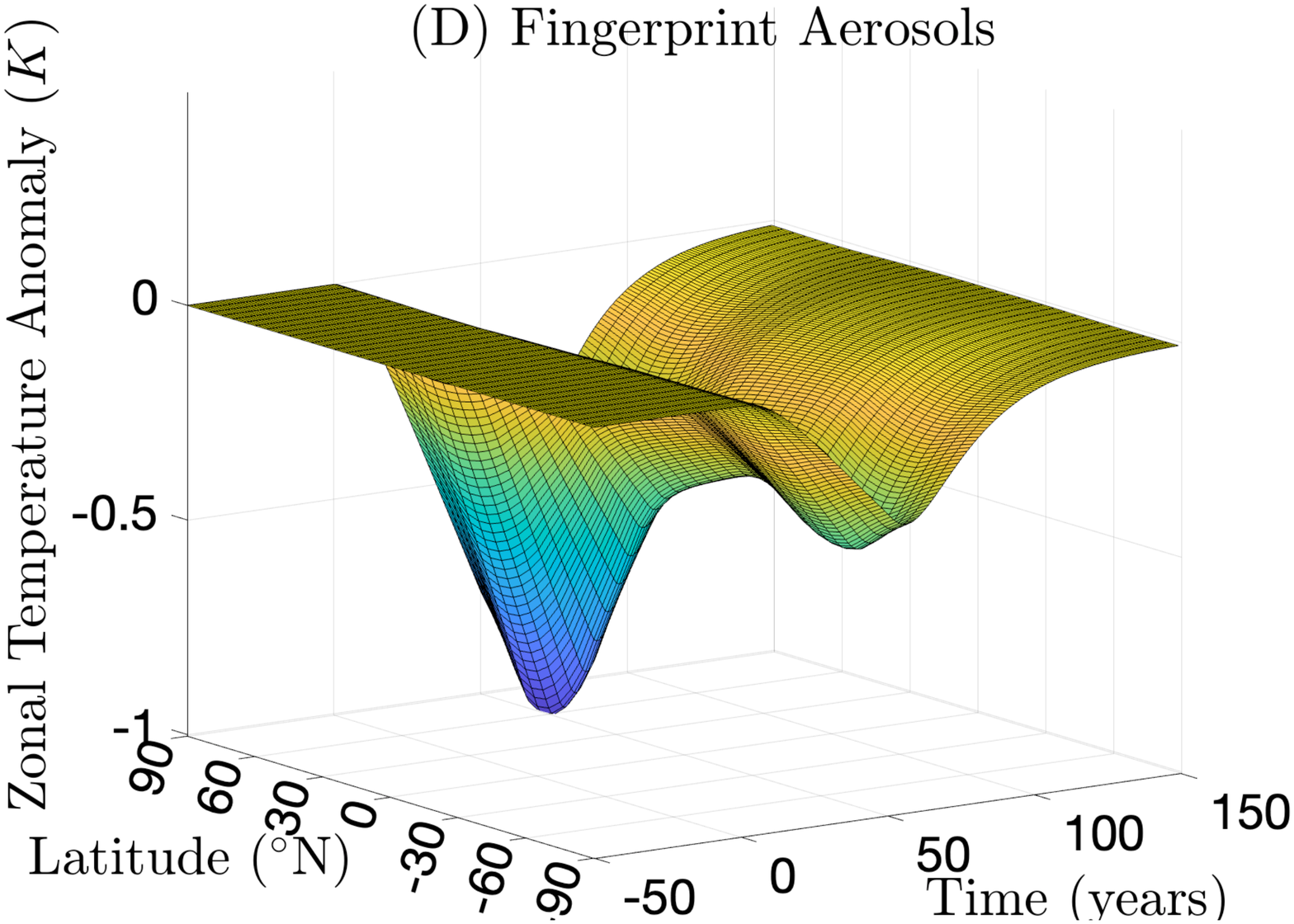}
\includegraphics[width=0.24\linewidth, height=.14\textwidth]{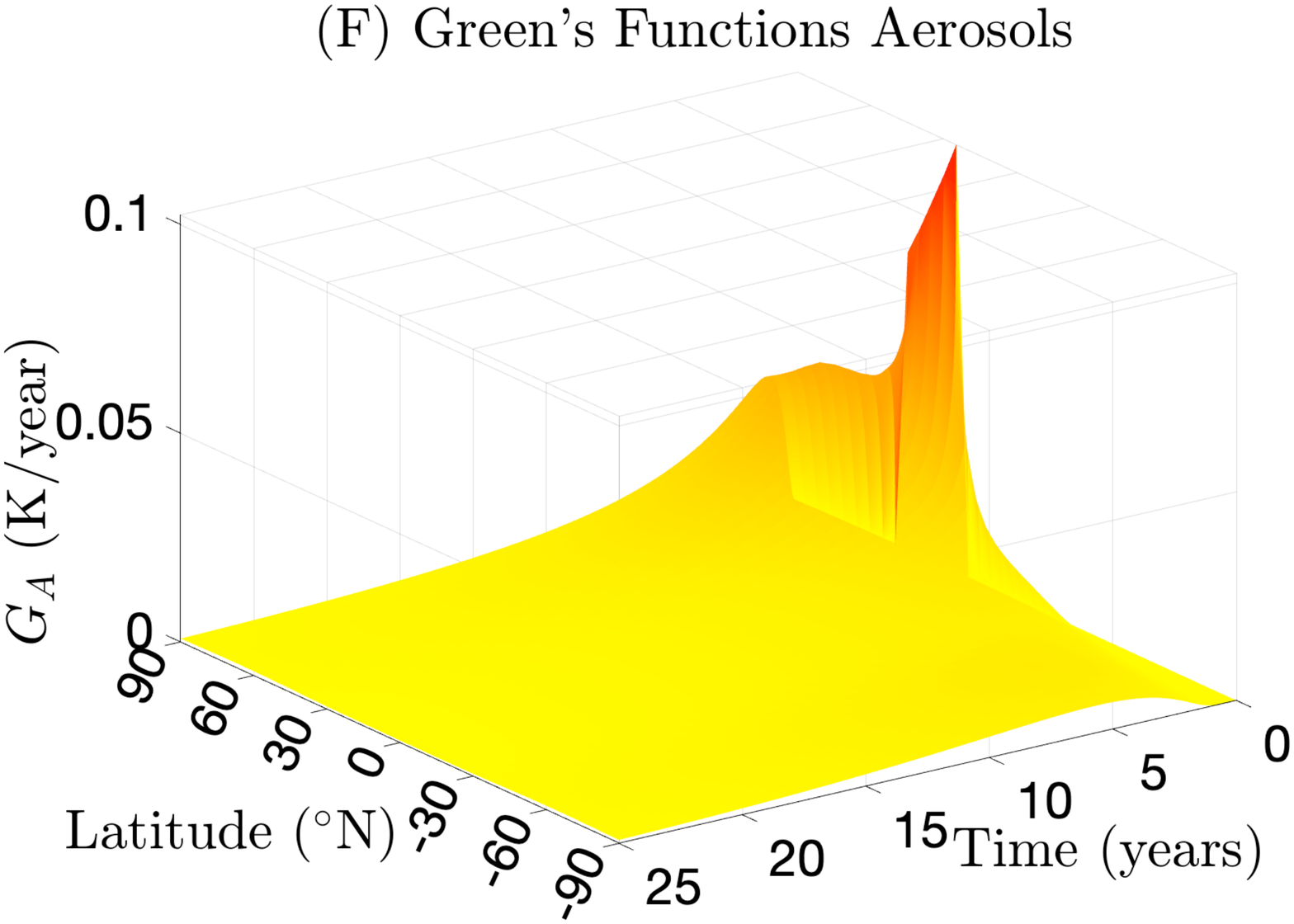}
\includegraphics[width=0.24\linewidth, height=.14\textwidth]{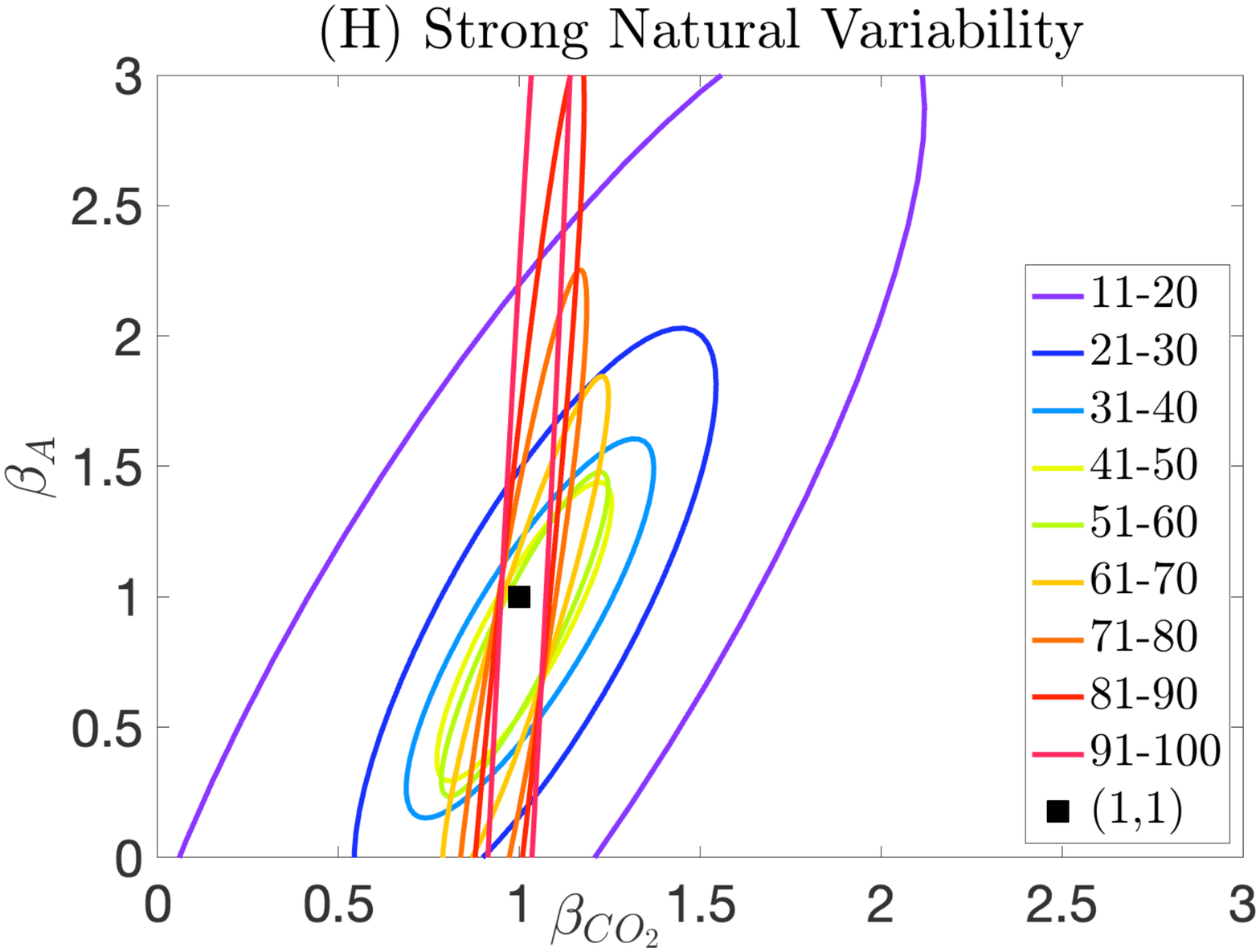}\\

\caption{{\bf OFM and link with response theory for the G-S model}. The climate change signal (from one ensemble member, in (A)) is decomposed according to Eq.~\eqref{eq:da3}: fingerprints associated with the $CO_2$ and aerosol forcings (shown in (C) and (D)) plus natural variability shown in (B).  The fingerprints shown in (C) and (D) are computed via response theory using the Green's functions (E) and (F)  in Eq.~\eqref{eq:da2}, respectively. The corresponding $\beta$-coefficients in Eq.~\eqref{eq:da3} and 95\% confidence regions are computed per decade, for each ensemble member,  in the cases of a weaker (G) and stronger (H) imposed natural variability. See text and Appendix B for details.} \label{Fig_EBM}
\end{figure*}
\emph{Energy Balance Model.}---The G-S model \cite{Ghil1976} can be seen as the ``hydrogen atom model'' of the climate system and 
 is a one-dimensional space-time reaction-diffusion model.  It describes the evolution of surface temperature  $T_S$ across different latitudes. To incorporate natural variability, we force the model with white noise. Details on the model and the D\&A experiments are provided in Appendix B.

The model is discretized in space,
 evaluating  $T_S$ at $d = 37$ latitudes. It is also subjected to $M = 2$ forcings. The first forcing mimics a $[CO_2]$ increase, ramping from a reference value to a final value over 100 years. The second forcing represents a localized increase in atmospheric aerosols, peaking around 50 years in the low-mid latitudes of the Northern Hemisphere and inducing a net cooling effect.

The climate change signal is the decadal $S=d=37$-dimensional vector of $T_S$  anomalies (across latitudes)  for ten decades following the start of the forcing. Along the lines discussed in Appendix D, we compute the Green's function for $T_S$ at each latitude for both forcings; see Figs.~\ref{Fig_EBM}e)-f). We use the Green's functions to compute the fingerprints via Eq.~\eqref{eq:da2}; see Figs.~\ref{Fig_EBM}c)-d). Following Eq.~\ref{eq:da3} we then apply the OFM for each individual forced run; see Fig.~\ref{Fig_EBM}a). We  derive for each decade estimates of the coefficients $\beta_{CO_2}$ and $\beta_{A}$ (for aerosols). We repeat the protocol for $N=100$  forced runs, thus obtaining confidence intervals for the  $\beta$'s, see Figs.~\ref{Fig_EBM}g)-h). 

 As the $CO_2$ and aerosols forcings strengthen over the first half of the century, the $\beta$'s 95\% confidence region shrinks. We obtain attribution to both forcings between years $30$ and $50$. However, as the aerosol signal weakens in the latter half of the century, attribution for the $CO_2$ forcing becomes clearer, while the  aerosol signal becomes indistinguishable from natural variability. This distinction arises because, despite their partially offsetting global effects, the $CO_2$ and aerosol forcings have different spatial patterns. Additionally, weaker natural variability simplifies the attribution process by reducing the noise level, as shown in Fig.~\ref{Fig_EBM}g). It is noteworthy that the confidence intervals for $\beta$ consistently center around 1.

\emph{PLASIM}---
We perform D\&A using the climate model PLASIM, which features $d=\mathcal{O}(10^5)$ degrees of freedom \cite{Fraedrich2005,Mehling2023}. We study the effect of $M=1$  forcing, namely the increase in $[CO_2]$. Along the lines of \cite{Ragone2016,Lucarini2017}, our protocol entails an annual 1\% increase of $[CO_2]$ from 360 ppm up to doubling, whilst afterwards $[CO_2]$ is kept constant. We run an ensemble of $N=40$ forced simulations.

The climate change signals of interest are the decadal averages of the  $T_S$ evaluated at the $S=64\times32=2048$  gridpoints. We use LRT to compute the fingerprints. We then  derive an estimate of $\beta_{CO_2}$ for each of the forced runs, thus deriving the $\beta_{CO_2}$ 95\% confidence interval;  see Appendices C and D.  Figure \ref{Fig_PLASIM}a) shows that as the global warming progresses, the $\beta$ confidence interval shrinks while being always centred around 1. The attribution of the anomaly signal to the $CO_2$ forcing is statistically robust already after the second decade.  LRT-based D\&A is effective  because LRT-based climate change projections are accurate: Fig.~\ref{Fig_PLASIM}b) shows the decadal  $T_S$ anomaly one century after the start of the forcing computed via LRT, while Fig.~\ref{Fig_PLASIM}c) shows the (modest) bias with respect to the corresponding ensemble average of the forced runs.
 \begin{figure*}[htbp]
\centering
\includegraphics[width=.32\textwidth,height=.18\textwidth]{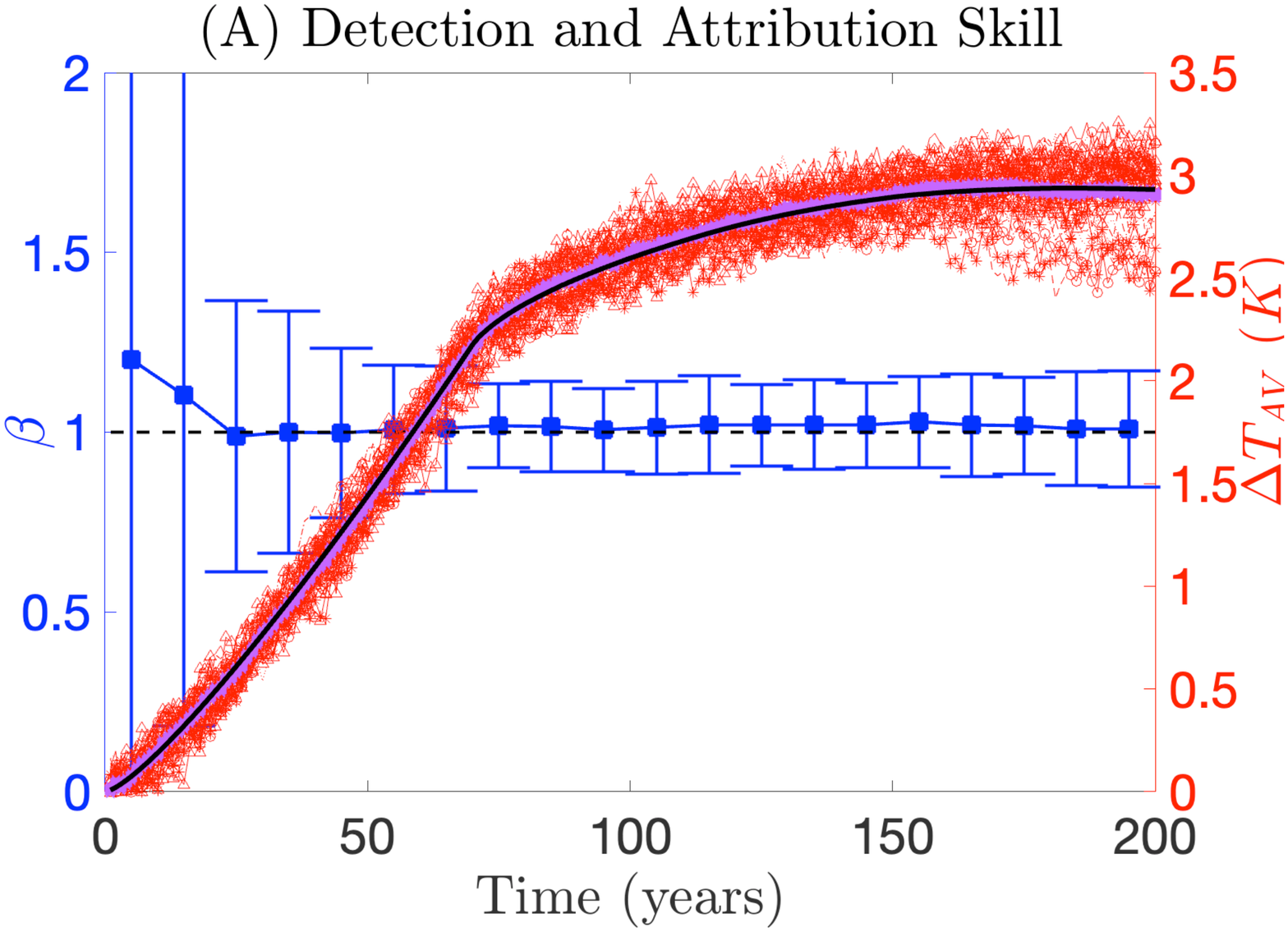}
\includegraphics[trim={2cm 3cm 2cm 1cm}, clip, width=.32\textwidth,height=.18\textwidth]{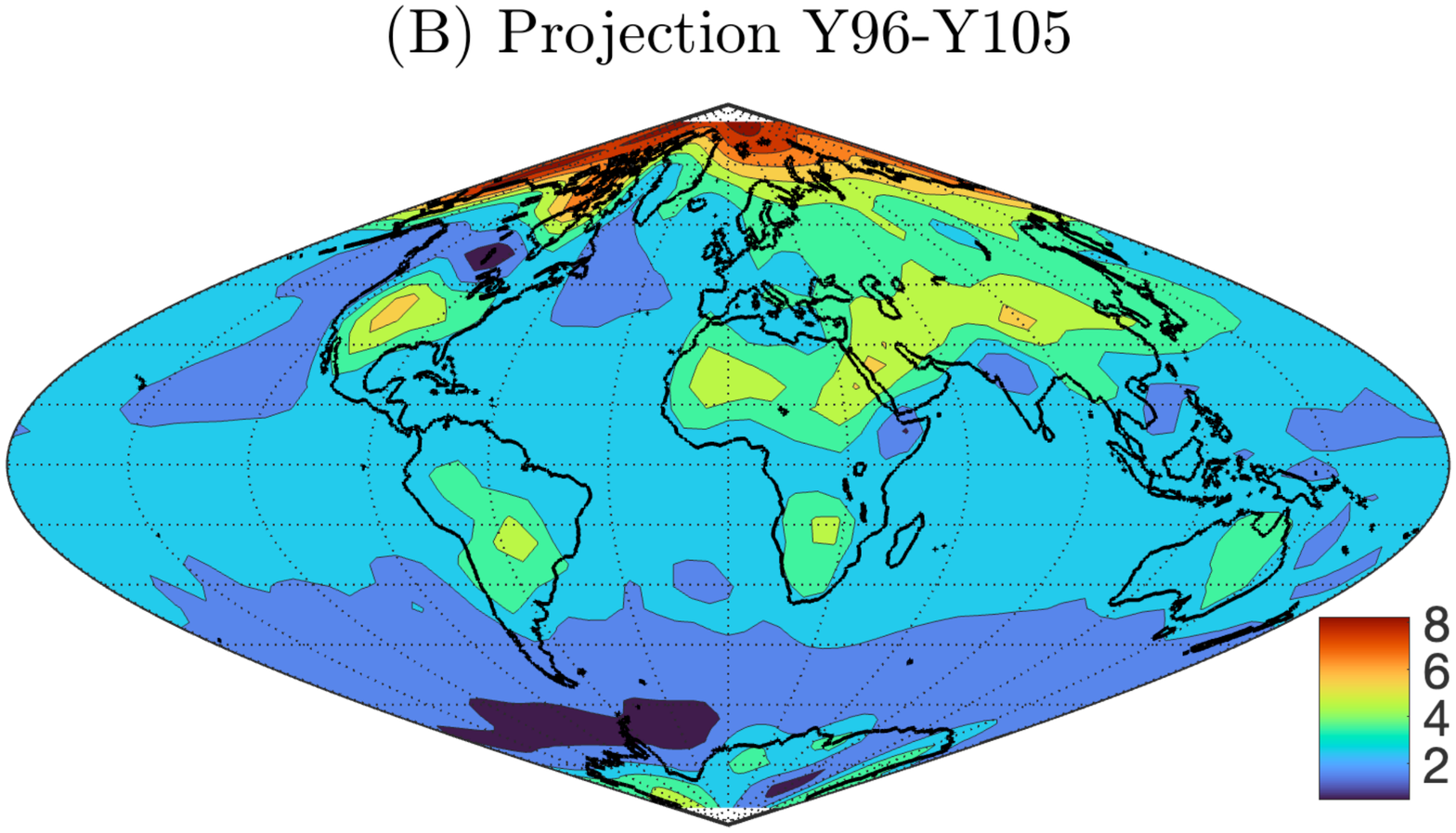}
\includegraphics[trim={2cm 3cm 2cm 1cm}, clip, width=.32\textwidth,height=.18\textwidth]{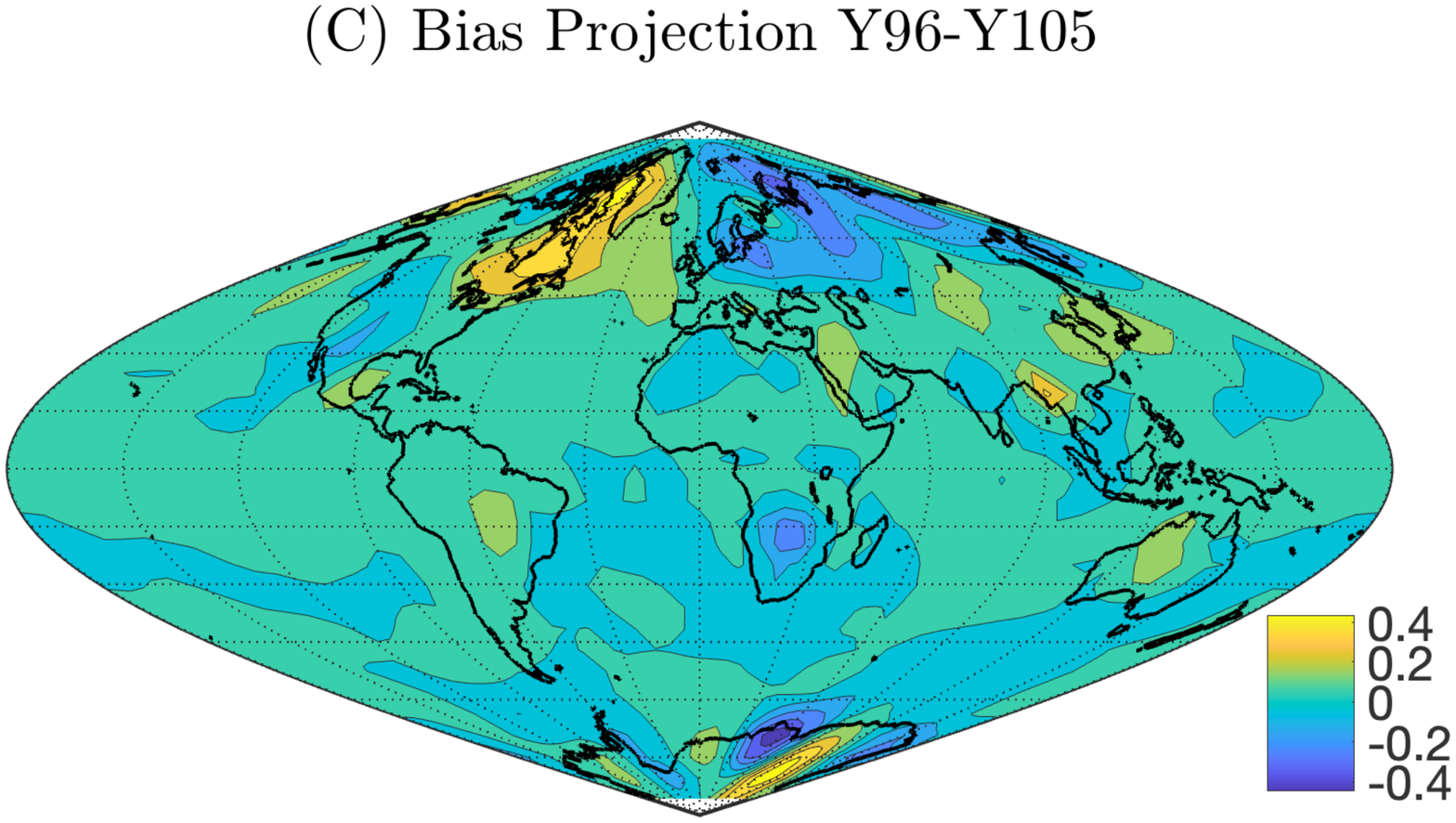}
\caption{{\bf D\&A of climate change for PLASIM forced experiments}. (A) 95\% Confidence interval for the $\beta$  factor (blue) of the $CO_2$ forcing fingerprint computed using  Green's functions. The uncertainty decreases as the signal-to-noise ratio increases, as shown by the globally averaged $T_S$ change $\Delta T_{AV}$  (red), ensemble average (thick magenta line; the black line shows the prediction) and  ensemble members (thin red lines). (B) Decadal average of the $\delta^{(1)} [T_S]$  (in $K$) field projection (Eq.~\eqref{eq:linear response time dependent}) during year 96 to 105  following the start of the forcing.  (C) Difference (in $K$) between the $\langle T_S \rangle$ anomaly of the forced runs and (B) for the year 96-105 average. The coast line is only indicative as it is at higher resolution than the land-sea mask used by the model. See text and Appendix C for details. }\label{Fig_PLASIM}
\end{figure*}
\vspace{-0.5cm}
\subsection*{\normalsize Nonlinear Fingerprints}
\vspace{-0.3cm}
\noindent The standard OFM relies  on the hypothesis of linearity of the response. Response theory can be extended to higher order terms \cite{ruelle_nonequilibrium_1998,lucarini2009b}.
The second order correction to the  expected value of a general observable $\Psi$ can be written as \cite{lucarini2008,LucariniColangeli2012}:
\begin{equation} \begin{aligned}\label{Eq_nln_response}
\delta^{(2)}[\Psi] (t) \hspace{-.1cm}= \hspace{-.2cm}\sum_{p,q=1}^M\epsilon_p\epsilon_q \hspace{-.1cm}& \int_\infty^t  \int_\infty^t \hspace{-.2cm} \d\tau_1 \d\tau_2  \GGG^{p,q}_{\Psi}(t-\tau_1,t-\tau_2)\\&
\times g_p(\tau_1)g_q(\tau_2),
\end{aligned}
\end{equation} 
where $\GGG_\Psi^{p,q}(t_1,t_2)$ is the second-order Green's function (causal in both time arguments) describing the joint effect of the $p^{th}$ and the $q^{th}$ forcing. 
One then has $\delta [\Psi](t) = \delta^{(1)}[\Psi] (t) +\delta^{(2)}[\Psi] (t)  +h.o.t,$, where $h.o.t.$ accounts for $o(\epsilon^2)$ terms. 
By accounting for $\delta^{(2)}[\Psi] (t) $ into the 
ensemble fluctuations $\langle \Psi \rangle_{\rho_\varepsilon^t} - \langle \Psi_k\rangle_0$
in Eq.~\eqref{Eq_fluc_ensemble}, we generalize Eq.~\eqref{eq:da2} as follows:
\be\label{eq:da2b}
Y_k(t)=\sum_{p=1}^M\tilde{X}_k^p(t) +\sum_{\ell=1}^{M^2}\tilde{Z}_k^{\ell}(t)+\mathcal{R}_k(t), \;\; k=1,\ldots,N.
\ee
where $\tilde{Z}_k^{\ell}(t)=\epsilon_p\epsilon_q \int \d\tau_1 \d\tau_2  \GGG^{p,q}_{\Psi}(t-\tau_1,t-\tau_2) g_p(\tau_1)g_q(\tau_2)$ with $ \ell=p+M(q-1)$. Higher-oder contributions can be constructed in a similar fashion. Indeed, Eq.~\eqref{Eq_nln_response} and its generalizations can be viewed as Volterra integrals where the Green's functions are interpreted as Volterra kernels \cite{Franz2006,Orcioni2014}.  

We then propose a generalization of the OFM able to deal with 
interacting responses 
by seeking a \textit{linear} regression for: 
\begin{equation} \begin{aligned}
\label{eq:danonlin}
Y_k&=\sum_{p=1}^{M}\tilde{X}_k^p\beta_p+\sum_{\ell=1}^{M^2}\tilde{Z}_k^{\ell}\gamma_{\ell}+\mathcal{R}_k, \quad k=1,\ldots,N\\
\tilde{X}_k^p&={X}_k^p+\mathcal{Q}_k^{p}, \quad \tilde{Z}_k^{\ell}={Z}_k^p + \mathcal{P}_k^{\ell}.
\end{aligned}\end{equation} 
The  factors $\beta$'s and $\gamma$'s  are unitary in the perfect model scenario, see Eq.~\eqref{Eq_nln_response}. The extra $M^2$ fingerprints compared to  Eq.~\eqref{eq:da3} can be constructed from selected climate model runs by suitable variations of the $\epsilon_p$'s \cite{lucarini2009b,gritsun2017}.  
A data-driven route leads to constructing the nonlinear response operator as feedforward neural networks \cite{Bodai2023}. After  training, the kernels can be computed from the weights and biases of the network \cite{Wray1994}.
 \vspace{-.6cm}
\subsection*{\normalsize Fingerprinting near Tipping Points}
\vspace{-.3cm}
\noindent 
Using the formalism of the Kolmogorov modes \cite{Chekroun_al_RP2}, which are the stochastic counterparts of the Koopman modes \cite{Budinisic2012,Kutz2016,Froyland2021}, we  write $\GGG^p_{\Psi}$ as a sum of terms, each  linked with a specific mode of natural climate variability. 
This expansion clarifies the link between free  and forced climate variability \cite{Leith75,Cortietal1999}: 
\begin{equation}\label{GreenH}
\GGG_{\Psi}^{p}(t) \approx \sum_{j=1}^{V}\sum_{\ell=0}^{m_j-1} \alpha^{\ell,p}_j(\Psi)\frac{1}{\ell!} e^{\lambda_jt}t^{\ell}, \quad t\geq 0, 
\end{equation}
with $\GGG_{\Psi}^{p}(t)=0$ when $t\leq 0$. This formula follows from the expansion of (temporal) correlations in terms of Kolmogorov modes, and relies on  neglecting the continuous spectrum \cite{Chekroun_al_RP2,Santos2022,LC2023}. Here $V$ is in general infinite and $m_j$  is the multiplicity of the $j^{th}$ eigenmode $\varphi_j$. These  are the eigenfunctions of the Kolmogorov operator $\mathcal{K}_{\Sigma}$  associated with Eq.~\eqref{Eq_unpertSDE}  (see Eq.~\eqref{Eq_Kolmo} in Appendix A. 
 The $\lambda_j$ are the corresponding eigenvalues, also known as Ruelle-Pollicott resonances \cite{Eckmann1985}. Often a finite number of these modes are approximated from time series in a reduced state space using Markov approximations \cite{Chek_al14_RP,TantetJSPIII} or delayed-embedding techniques \cite{Froyland2021}.  
 
 They encapsulate the decay rate and frequency of oscillation of the natural modes of variability (Sec.~2.3 in \cite{Chekroun_al_RP2}), whilst the factors $\alpha^{\ell,p}_j$ depend on the applied forcing and choice of observable \cite{LC2023}. The spectral gap $\gamma=\mathfrak{Re}(\lambda_1)$  associated with the slowest decaying mode $\varphi_1$
indicates the proximity to tipping. Rough dependence of statistics on parameters is found as $\gamma \rightarrow 0$ \cite{Chek_al14_RP}. At a tipping point, $\gamma \rightarrow 0$,  and Eq.~\eqref{GreenH} indicates that 
any Green's function decays sub-exponentially, irrespectively of the observable and forcing, unless the corresponding $\alpha$-coefficient vanishes.
At criticality the response may  diverge \cite{Chek_al14_RP,Tantet2018,AshwinJSP,Santos2022}. Hence, if either the  climate system or the model used for fingerprinting is close to a tipping point, D\&A via OFM might experience major uncertainties and biases.

If each isolated eigenvalue has unitary multiplicity, the time-lagged correlation for $\Psi$ is $\mbox{Corr}_{\Psi}(t)=\sum_{j=1}^N a_j(\Psi) e^{\lambda_j |t|} $ with $a_j(\Psi)$ given in Corollary 1 of \cite{Chekroun_al_RP2}.
If $\gamma \rightarrow 0$, $\mbox{Corr}_{\Psi}(t)$ decays subexponentially when $\Psi$ has a non-vanishing projection onto $\varphi_1$. This is the so-called critical slowing down associated with tipping behaviour  \cite{scheffer2012anticipating,boers2021critical}, which was originally discovered for continuous phase transitions \cite{stanley1971}. If $\Psi\propto\varphi_1$, $\mbox{Corr}_{\Psi}(t)\propto e^{\lambda_1|t|}$  and $\varphi_1$ is the degenerate fingerprinting's critical mode \cite{Held2004},  encoding the ``natural'' tipping observable.
\vspace{-0.5cm}
\subsection*{\normalsize Discussion and Perspectives}
\vspace{-0.3cm}
\noindent 

OFM has been instrumental in shaping modern climate change science and has had significant societal impact \cite{IPCC13,IPCC2021}. However, recent criticisms highlight potential underestimates of uncertainties and even question the statistical basis of OFM \cite{Li2021,McKitrick2022,Chen2022}.
This work addresses this issue by deriving OFM equations for D\&A from LRT for nonequilibrium systems, thus providing a solid physical and dynamical foundation for OFM. In particular, 
the causality principle embodied in the Green's functions aligns closely with Pearl's interventionist angle  \cite{Pearl2009}, which is key for D\&A studies. Our derivation extends OFM applicability to virtually any complex system---ecosystems, quantitative social sciences, finance---where attributing observed changes to multiple forcings is desired. 

This study offers several key insights for fingerprint analysis in climate science as listed below.

\noindent{\it Systematic Attribution}:  In the perfect model scenario where all forcings are included, this work explains why LRT provides accurate attribution in linear regimes. It also clarifies why the weighting factors ($\beta's$) are equal to one in such a scenario. However, a crucial limitation is identified: mismatches between the model's and the climate system's natural variability patterns (Kolmogorov modes) can distort Green's functions, affecting fingerprint accuracy across different timescales. This is especially concerning near tipping points. Future research will focus on linking natural and forced variability through Kolmogorov mode analysis.

 \noindent{\it Combining Models}: Combining fingerprint estimates from different models can be problematic due to differences in their Kolmogorov modes, potentially leading to  errors, especially near tipping points. Instead, using an ensemble approach with a single model aligns with the mathematical framework.

  \noindent{\it Critical Modes and Early Warning Signals}: Since Kolmogorov modes reflect proximity to critical transitions, they offer a reliable basis for degenerate fingerprinting, which is key for defining early warning indicators for tipping points \cite{Held2004} and for defining the critical mode of variability.

          \noindent{\it Nonlinear extension to the OFM}: Nonlinear response theory allows for a more powerful Optimal Fingerprint Method (OFM) framework.
 Even with significant nonlinearities due to strong forcing interactions, the method remarkably leads to a linear regression problem. These nonlinearities manifest as additional fingerprints, providing practical benefits for climate research and other complex systems. Future work will explore these applications in detail.    
          
To demonstrate the theory's potential, we applied its key findings to perform D\&A of simulated climate change on two models: the simple yet historically relevant and physically informative one dimensional G-S energy balance model, and the more realistic three dimensional coupled climate model PLASIM. The results showcase how LRT and OFM work hand in hand. Green's functions enable computation of accurate fingerprint across timescales, while the signal-to-noise ratio heavily influences attribution robustness.
Our approach paves the way for applying LRT to compute fingerprints  in state-of-the-art ESMs \cite{Eyring2020}. The effectiveness of LRT for climate change projections in ESMs has already been established \cite{Lembo2020}.  Successfully implementing LRT in this way could significantly improve our understanding of the current climate crisis.

%
 %
 \vspace{-0.5cm}
\appendix
\subsection*{Appendix A: Green's Function and Response}
 \vspace{-0.3cm}
Near equilibrium, the standard form of the fluctuation-dissipation theorem (FDT) \cite{Kubo1966,abramov2007,pavliotisbook2014} allows for predicting forced fluctuations  in terms of readily accessible and intuitive correlations of observables in the unperturbed system.
This form of the FDT has been applied to predict climate models' response to changes in the solar irradiance \cite{North1993}, greenhouse gases concentration \cite{Cionni2004,Langen2005}, and to study the impact of localized heating anomalies \cite{gritsun2007}. Yet, this approach can lead to potentially large errors  \cite{gritsun2017}. 
One can compute the climate response via LRT, bypassing  the FDT, for models of extremely diverse complexity \cite{Lucarini2011,Ragone2016,Lucarini2017,Lembo2020,Torres2021a,Bastiaansen2021}. 
We  frame climate change by 
expanding the statistical state, $\rho_\epsilon^t$, and solving the time-dependent FPE associated with the  SDE \eqref{eq:sto ode 2} as:
\begin{equation} \begin{aligned}\label{eq:expansion time dependent density}
\rho_\epsilon^t(\xx) = \rho_0(\xx) + \sum_{p=1}^M\epsilon_p & \rho_{p}^t(\xx)
+ h.o.t.
\end{aligned}
\end{equation}
where $\rho_0(\xx)$, is the reference unperturbed climate described by the 
stationary probability density associated with 
Eq. \eqref{Eq_unpertSDE}. Such an  expansion is the starting point of virtually any linear response formula; see \cite{Lucarini2016,SantosJSP} for a discussion on the radius of convergence.
The expected value  of 
$\Psi$ at time $t$ is given by
\begin{equation} \begin{aligned}\label{eq:response function a}
	 \langle \Psi \rangle_{\rho_\epsilon^t}&=\int \d\xx\rho^t_\epsilon(\xx) \Psi(\xx)  \\
	&\approx \int  \d\xx \rho_0(\xx)  \Psi(\xx) + \sum_{p=1}^M\epsilon_p   \int   \d\xx \rho_{p}^t(\xx) \Psi(\xx).
\end{aligned}\end{equation}

Under natural assumptions valid for a wide class of stochastic systems \cite{Hairer2010},
the \emph{linear response} approximates the ensemble anomaly i.e.~$\langle \Psi \rangle_{\rho_\varepsilon^t} - \langle \Psi \rangle_0=\delta [\Psi](t)$  by $\delta^{(1)}[\Psi] (t)$ given by Eq.~\eqref{eq:linear response time dependent}. The Green's functions $\G^p_{\Psi}$ are \cite{Santos2022}:
\begin{equation} \label{eq:Green}
\G^p_{\Psi}(t)=\Theta(t) \hspace{-1ex}\int  \hspace{-1ex}  \d\xx \rho_0(\xx) \bigg(e^{t\mathcal{K}_{\Sigma}}
 \Psi(\xx)\big[\cL_p\log(\rho_0)\big] (\xx)\bigg),
\end{equation}
where $\Theta(t)$ is the Heaviside distribution ensuring causality  \cite{ruelle2009,Lucarini2017,Lucarini2018JSP}. 
The operators $\mathcal{K}_{\Sigma}$, and $\cL_{p}$ are \cite{Santos2022}:
\bea \label{Eq_Kolmo} 
\mathcal{K}_{\Sigma} \Psi&= \FF \cdot \nabla \Psi  + \frac{1}{2}\Sigma\Sigma^{T}: \nabla ^2  \Psi \\
\cL_p \rho &= -  \nabla \cdot\left( \GG_p  \rho\right), \hspace{0.2cm}p=1,\ldots,M.
\eea
In Eq.~\eqref{Eq_Kolmo}, $\mathcal{K}_{\Sigma}$ denotes the Kolmogorov operator associated with Eq.~\eqref{Eq_unpertSDE},  ":" denotes the Hadamard product, and $e^{t\mathcal{K}_{\Sigma}}$ in Eq.~\eqref{eq:Green} is the corresponding Markov semigroup  \cite{Chekroun_al_RP2}. Instead,  $\cL_p$ indicates the correction to the Kolmogorov operator due to the perturbation included in Eq.~\eqref{eq:sto ode 2}. These formulas  provide a general version of the FDT, as the Green's functions are lagged correlations of the observables  $\Phi=\cL_p \log (\rho_0)$ and $\Psi$. 

As made very clear in Ruelle's derivation of response formulas for nonequilibrium systems \cite{ruellegeneral1998}, Eq.~\eqref{eq:linear response time dependent} combined with the causality of $\G^p_{\Psi}$ implies that the effect of the perturbation comes strictly after the forcing is applied, thus defining a time-ordered causal link and flow of information. This is excellently aligned with the concept of interventions and counterfactual reality needed to define causality in Pearl's sense \cite{Pearl2009}.
 \vspace{-0.5cm}
\subsection*{Appendix B. The Ghil-Sellers Model}
 \vspace{-0.3cm}
The G-S energy balance model \cite{Ghil1976} captures the essence of Earth's system radiative budget. It accounts for the absorption and reflection of solar radiation, the emission of infrared radiation back to space, and the large-scale heat transport from  tropical regions towards the poles  \cite{Peixoto1992}. The G-S model has been instrumental for understanding why Earth's climate is metastable 
\cite{Bodai2015,Ghil2020} and describes the evolution of the zonally-averaged surface temperature $T_S(x,t)$ where $x=2\phi/\pi$ - the normalized latitude $\phi$ - lies in $[-1,1]$, and $t$ is time:
\bea\label{eq:pde}
 c(x) &\partial_t T_S  =\left(\frac{2}{\pi}\right)^2\frac{1}{\cos\left(\frac{\pi x}{2}\right)}\partial_x\left(\cos\left(\frac{\pi x}{2}\right)\kappa(x,T_S)\partial_xT_S\right) \\
 + &\mu Q(x)\left(1 - \alpha(x,T_S)\right)  - \sigma T_S^4\left(1 - m\tanh(c_3T_S^6)\right), 
\eea
with von Neumann boundary conditions  
$\partial_xT_S(-1,t) = \partial_xT_S(1,t) = 0$. Here $c(x)$ is the effective heat capacity of the atmosphere, land, and ocean per unit surface area at $x$. The first term on the right hand side (RHS) represents the meridional heat transport as a diffusive law, where $\kappa(x,T_S)$ incorporates the effects of sensible and latent heat transport. %
The solar forcing (second term on the RHS) is modulated by the solar constant $\mu$, the  irradiance $Q$ and the albedo $\alpha$. The longwave emission (third term on the RHS) follows the Stefan-Boltzmann's law, modified by the greenhouse effect, whose intensity is modulated by the constant $m$ (Earth's grayness). 
Further details on the G-S model are reported in \cite{Bodai2015}, which is our reference for all constants and tabulated functions.  

\emph{Numerics.}---We implement the G-S model  by discretizing the latitude in $d=37$ increments of 5$^\circ$ and by using a time step of one day. We discretize the spatial derivative operators  via standard centered differences. To mimic the effects of the unresolved degrees of freedom and account for  natural variability, we add a white noise forcing to Eq.~\eqref{eq:pde}, with spatial correlation matrix given by $\eta_0\mathbf{I}_d$ with $\eta_0\geq0$. We use the Euler-Maruyama scheme for time integration.  We  choose as reference state the warm climate established with the present-day solar constant $\mu=1$ and $\eta_0=0$ (see Fig. 1a in \cite{Bodai2015}. Our ensemble runs are performed using $\eta_0=0.2$ (reference, strong noise case) and $\eta_0=0.1$ (weak noise case).

\emph{Climate Change Experiments.}---We simultaneously perform a) an increase in $m$, which mimics the increase in $[CO_2]$; and b) a reduction of the incoming radiation in the region $[25^\circ N,45^\circ N]$, mimicking the effect of aerosol injection in the atmosphere in the low-to-mid latitudes of the Northern Hemisphere. The forcing a) is realized by increasing linearly over 100 years $m$ in Eq.~\eqref{eq:pde} from  $m=m_0=0.5$ to $m=m_0+\delta m$, where $\delta m=0.01$, and keeping the reached value of $m$ constant afterwards. This protocol corresponds roughly to a classical IPCC $[CO_2]$ stabilization scenario. The forcing b) is realized by multiplying in the region $[25^\circ N,45^\circ N]$  $\mu$ times a factor $\nu$ that decreases linearly in 50 years from  $\nu_0=1$ to $\nu=1-\delta\nu$, $\delta \nu=0.012$, and then letting the perturbation fade away with an exponential law with decay time of 20 years. This corresponds to a typical IPCC technology transition scenario, where aerosol emissions first peak and then decay \cite{IPCC13}.

\vspace{-0.5cm}
\subsection{Appendix C: The Climate Model PLASIM-LSG}
 \label{sec:PlaSim} \vspace{-0.3cm}
The Planet Simulator (PLASIM) is an open-source low-resolution climate model with a graphical user interface and comprehensive diagnostics suites \cite{Fraedrich2005,Lembo2019}. 
PLASIM has been widely used to simulate  
 past climate conditions \cite{Lucarini2010a,Roscher2011}, to test specific atmospheric processes \cite{Hertwig2014}, to look into the climate  via nonequilibrium thermodynamics \cite{Lucarini2010b, Fraedrich2008}, and to study circulation regimes of interest for exoplanetary research \cite{Boschi2013,Paradise2021}. PLASIM has  also been used as a testbed to investigate the climate with statistical mechanical tools. I has been shown that one can compute climate change projections via response theory 
\cite{Ragone2016,Lucarini2017}, study  tipping points through a functional analytical angle \cite{Tantet2018b}, investigate weather extremes using rare event algorithms \cite{Ragone2018,Wouters2023}, and analyze the climate metastability using field-theoretical approaches \cite{Margazoglou2021}. 

The atmospheric dynamics is modelled using the multi-layer primitive equations, whilst 
moisture is included by 
transport of water vapour. We use a T21 spectral resolution, 
corresponding to $\approx5.6^\circ$ resolution in latitude and longitude), and 10 vertical levels. The land surface scheme uses five diffusive 
layers for the temperature and a bucket model for the soil hydrology. 
Here, following \cite{Mehling2023}, we consider as fully coupled ocean component the Large Scale Geostrophic (LSG) circulation model, which is based on the primitive equations and on the thermodynamics of saline water. 
 The version used here 
 has an effective  $\approx3.5^\circ$ resolution in latitude and longitude and 22 vertical layers. The  code is available here:  {\small \url{https://github.com/jhardenberg/PLASIM}.} The model is supplemented by well-tested parametrizations for unresolved processes, undergoes periodic forcing due to the orbital annual cycle and has homogeneous atmospheric $[CO_2]$, which can  be changed following a chosen temporal protocol, as done here.
 
 \begin{figure}[htbp]
\centering
\includegraphics[width=.395\textwidth]{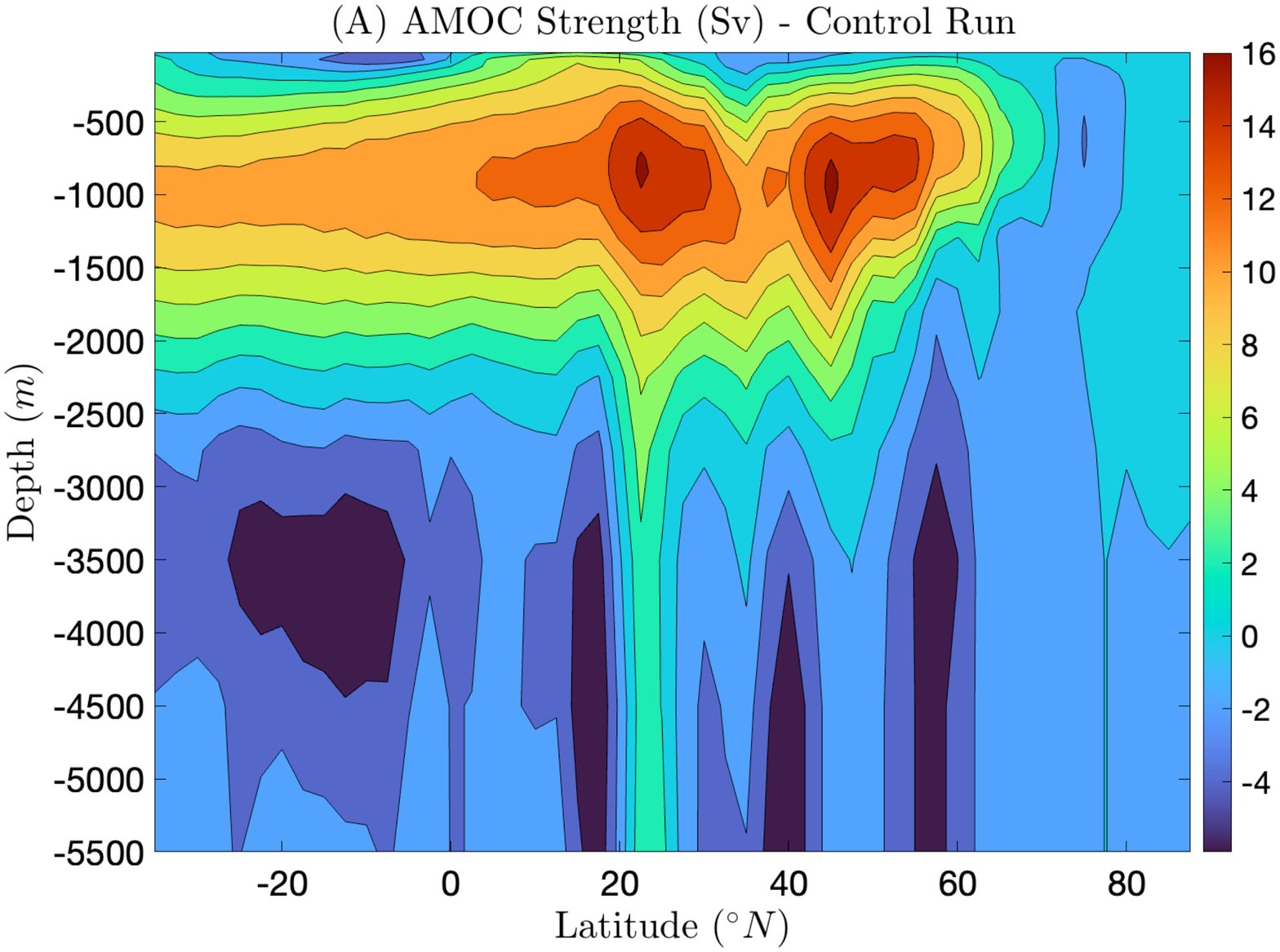}
\includegraphics[width=.395\textwidth]{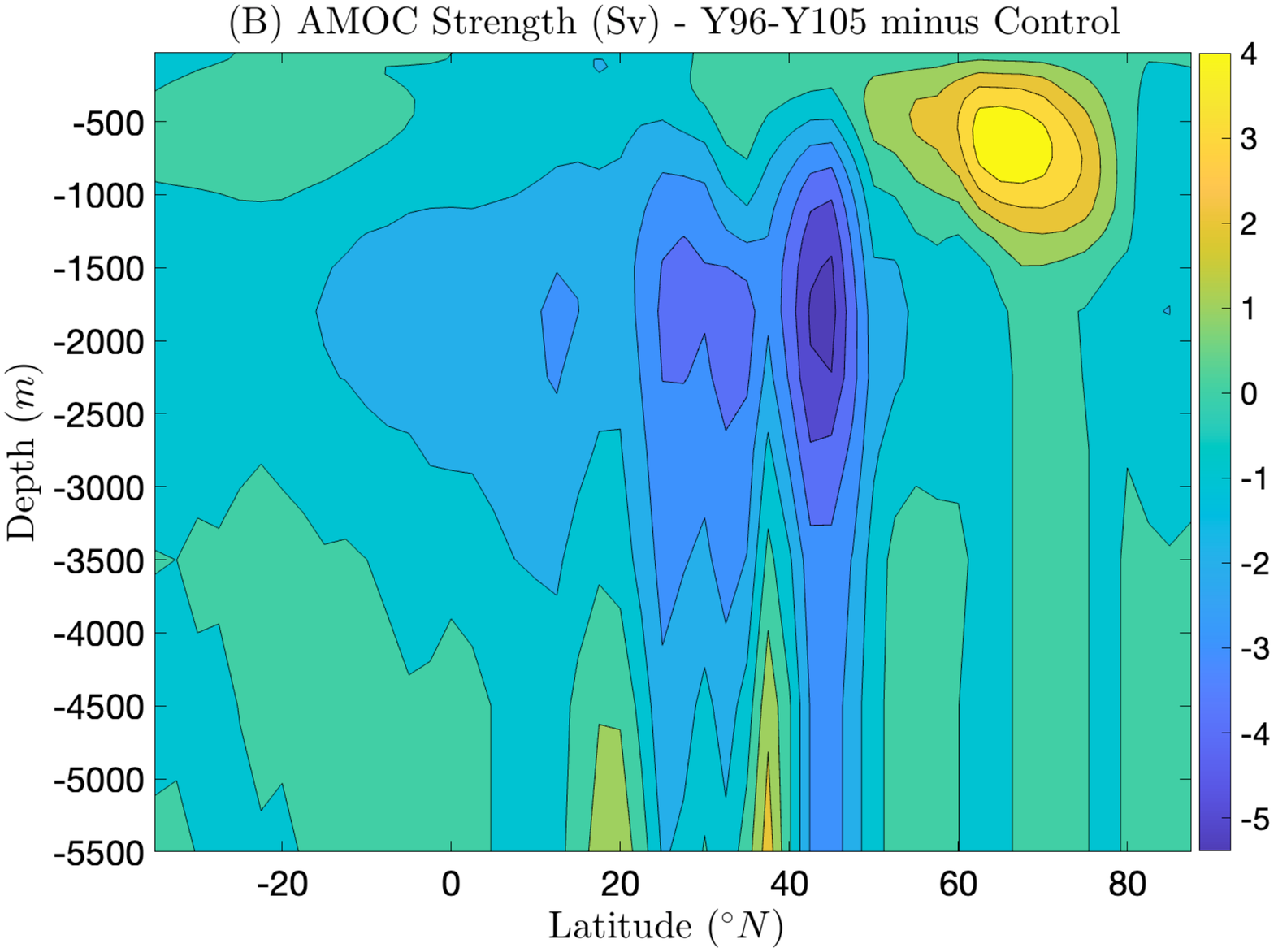}
\caption{PLASIM AMOC streamfunction showing northward transport at surface and return flow in depth (in $Sv=10^6m^3s^{-1}$). (A) Climatology for reference $[CO_2]$. (B) Ensemble mean of the decadal anomaly  30 years after the end of the  $1\%$ yearly $[CO_2]$ increase ramp.} \label{AMOC} 
\end{figure}
\emph{Climate Change Experiments.}---We consider a 
1\% increase of $[CO_2]$ starting from a reference value of 360 ppm , with the system at steady state, up to doubling (which occurs after $\approx$ 70 years). $[CO_2]$ is then kept constant at 720 ppm until the system reaches  a newly established steady state. The initial conditions for the $N=40$ climate change runs are chosen from a 4000-year long control run with reference $[CO_2]$, with the $j^{th}$ ensemble member being initialised on the first day of  year $1 + 100(j-1)$ of the control run. Whilst $[CO_2]$ grows exponentially up to doubling, the actual radiative forcing acting on the system grows linearly with time, see \cite{Ragone2016,Lucarini2017}.  

Despite its relative simplicity, PLASIM has a fairly realistic response to $CO_2$ forcing, featuring a climate sensitivity \cite{vonderHeydt2016} of $\approx 3$ $K$ and an amplified warming over land in the northern Hemisphere and in the high-latitude belt (polar amplification). A strong  reduction of the warming occurs in the North Atlantic, see Fig. \ref{Fig_PLASIM}b). The so-called \textit{warming hole} or \textit{cold blob} \cite{Drijfhout2012,Bryden2020,Lembo2020} is due to the weakening of the Atlantic meridional overturning circulation (AMOC);  see Fig.~\ref{AMOC}. The AMOC is responsible for the ocean meridional heat transport \cite{Kuhlbrodt2007} and is a tipping element \cite{Lenton.tip.08} that is  approaching criticality  \cite{Ditlevsen2023,Westen2024}. 

 \vspace{-0.5cm}
\subsection{Appendix D: Computing the Green's functions} \vspace{-0.3cm}
Both for  the G-S and the PLASIM-LSG models, we follow the strategy proposed in \cite{Ragone2016,Lucarini2017} for estimating the Green's functions $\mathcal{G}^p_\Psi$. 
We first select $M$ statistically independent initial conditions drawn from the steady-state unperturbed system. 
In each run, we choose as time modulation of $p^{th}$ forcing in Eq.~\eqref{eq:sto ode 2}, $g^p(t)=\Theta(t)$, which amounts to an instantaneous switch of this forcing only. 
By taking the average over the $M$ runs, we estimate $\delta^{(1)}[\Psi] (t)$, and - see Eq.~\eqref{eq:linear response time dependent} - we derive  $\mathcal{G}^p_\Psi (t) \approx 1/{\epsilon_p} \mathrm{d}(\delta^{(1)}[\Psi] (t))/\mathrm{d}t$. 

For the G-S model, the Green's functions describing the $T_S$ response at the $37$ latitudes due to changing $m$ are computed by 
instantaneously increasing  $m=m_0\rightarrow m_0+\delta m$. To obtain the Green's functions for the second forcing, 
we  instantaneously apply $\nu=\nu_0=1\rightarrow1-\delta \nu$. 
For PLASIM-LSG, we compute the Green's functions for the $T_S$ anomalies at the $2048$ gridpoints due to $[CO_2]$ forcing by instantaneously doubling  $[CO_2]$ from 360 ppm to 720 ppm; see also \cite{Ragone2016,Lucarini2017,Lembo2020}.

\begin{acknowledgments}
\noindent{\bf Acknowledgements}\\
{\bf Funding:} VL acknowledges the support by the EU Horizon 2020 project TiPES (grant no.~820970), the EU Horizon Europe project ClimTIP (grant no. 101137601),  by the Marie Curie ITN CriticalEarth (grant no.~956170), and by the EPSRC project LINK (Grant No. EP/Y026675/1). MDC acknowledges the European Research Council  under the EU Horizon 2020 research and innovation program (grant no.~810370). This work has been also supported by the Office of Naval Research (ONR) Multidisciplinary University Research Initiative (MURI) grant N00014-20-1-2023, and by the National Science Foundation grant DMS-2407484. The authors wish to thank Reyk B\"orner for support on the numerical simulations performed with PLASIM and two anonymous reviewers for providing very useful constructive criticism and insightful comments.

{\bf Author contributions:} 
This work initiated from discussions between VL and MDC.
Both authors conceptualized the approach and the mathematical details.
VL performed the numerical calculations  behind Fig.~\ref{Fig_EBM} and performed the data analysis leading to Figs.~\ref{Fig_PLASIM}-\ref{AMOC}. VL and MDC interpreted the main results and wrote the manuscript.
{\bf Competing Interests:} The authors declare that they have no competing interests.
{\bf Data and materials availability:} All data needed to evaluate the conclusions of this work are present in the paper.
\end{acknowledgments}

\end{document}